\documentclass {article}
\usepackage{amsfonts}
\usepackage{graphicx}
\DeclareSymbolFont{AMSb}{U}{msb}{m}{n}
\DeclareSymbolFontAlphabet{\mathbb}{AMSb}
\newcommand{\complex}{\kern.1em{\raise.47ex\hbox{
            $\scriptscriptstyle |$}}\kern-.40em{\rm C}}

\newcommand{\Partial}[2]{\frac{\partial #1}{\partial #2}}

\title{Colored Hofstadter butterflies}
\begin{document}

\author{J.~E.~Avron \\ Department of Physics, Technion, 32000 Haifa, Israel}
\maketitle
\begin{abstract}
I explain  the thermodynamic significance, the duality and open
problems associated with the two colored butterflies shown in
figures \ref{fig:tb} and \ref{fig:sl}.
\end{abstract}

\section{ Overview}
My aim is to explain what is known about the thermodynamic
significance of the two colored butterflies shown in figures
\ref{fig:tb} and \ref{fig:sl} and what remains open. Both diagrams
were made by my student, D. Osadchy \cite{ref:osadchy}, as part of
his M.Sc.\ thesis. I shall explain their interpretation as the
$T=0$ phase diagrams of a two dimensional gas of charged, though
non-interacting, fermions. Fig.~\ref{fig:tb} is associated with
weak magnetic fields (and strong periodic potentials)  while
Fig.~\ref{fig:sl} with strong magnetic fields (and weak periodic
potentials). The two cases are related by duality. The duality,
which is further discussed below, is manifest if colors are
disregarded.

The horizontal coordinate in both figures is the chemical
potential $\mu$ and the vertical coordinate is  proportional to
the magnetic induction $B$ in fig.~\ref{fig:tb} and $1/B$ in
fig.~\ref{fig:sl}.  The colors represent the quantized values of
the Hall conductance, i.e. represent integers \footnote{ The
quantum unit of conductance, $e^2/h$, is $1/2\pi$, in natural
units where $e=\hbar=1$.}. Warm colors represent positive
multiples and cold colors represent negative ones: Orange
represents 2, red 1, white 0, blue $-1$ etc.

{\bf Remark:}{\quad \it It is problematic to  represent integers
by colors with good contrast between nearby integers. This is
related to the fact that colors are not ordered on the line but
rather are represented by the simplex $(r,g,b)$ with $r+g+b=1$.
(Pure colors are located on the boundary of the simplex). The
assignment in the figures becomes problematic for large, positive
or negative, integers: Large positive integers are not represented
anymore by warm colors but rather by yellow and green. }

I shall also present an open problem. Namely, how do these
diagrams change if one replaces the magnetic induction $B$ by the
magnetic field $H$ as the thermodynamic coordinate.
\section{Some history}

That the Hall conductance took different signs in different metals
was  an embarrassment to Sommerfeld theory. Since charge is
carried by the electrons one sign was predicted.  The wrong sign
was called the anomalous Hall effect and was explained by R.
Peierls \cite{ref:peierls} who showed that the periodicity of the
electron dispersion $\epsilon(k)$ plus the Pauli principle allow
for either sign, depending on $\mu$. This subsequently lead to the
important concept of holes as charge carriers---a term not used by
Peierls in his original work.

The electron-hole anti-symmetry of the Hall conductance is seen in
Fig.~\ref{fig:tb} where cold and warm colors are interchanged upon
reflection about the vertical axis. However, the figure is much
more complicated than what Peierls had in mind.


Mark Azbel \cite{ref:azbel} realized that the Schr\"odinger
equation in a periodic potential and magnetic field had
tantalizing spectral properties. But it was the graphic rendering
of the spectrum by D. Hofstadter \cite{ref:hofstadter}, shown in
Fig.~\ref{fig:hof}, (and his scaling rules,) that brought the
problem into limelights. The richness of spectral properties is a
result of competing area scales: One dictated by the unit cell of
the underlying periodic potential and the other by the area that
carries one unit of magnetic flux. When $\Phi$ is rational the two
areas are commensurate, when it is irrational, they are not. At
$T=0$ the electrons gas is coherent on large distance scales and
commensuration lead to interference phenomena that affect spectral
properties at very small energy scales. The delicate spectral
properties attracted considerable attention of a community of
spectral analysts. Reference \cite{ref:avila} is a pointer to a
rich and wonderful literature on the subject.


In a seminal work, TKNN \cite{ref:tknn} realized that the Hall
conductance of the Hofstadter model admits a topological
characterization in terms of Chern numbers.  This discovery has an
an interesting piece of lost history.  In fact S. Novikov was
apparently the first to realized the topological significance of
the spectral gaps for Bloch electrons in magnetic fields
\cite{ref:novikov}. However, he missed their significance as Hall
conductance. TKNN \cite{ref:tknn} were not aware of the work of
Novikov. Instead, they were motivated by a puzzle that follow from
applying the Laughlin argument for the quantization of the Hall
conductance to the Hofstadter model. By essentially reinventing
the proof of integrality of Chern numbers in a special case, they
showed that  whenever the Fermi energy is in a gap the Hall
conductance is quantized.

In the two diagrams, figs. \ref{fig:tb},\ref{fig:sl}, the Hall
conductance is quantized almost everywhere. The set of points
where the Hall conductance is not quantized is a set of zero
measure and so invisible. This is related to the fact that the
spectrum is a set of measure zero (see e.g. \cite{ref:avila} and
references therein).

\section{Thermodynamics considerations}

It is interesting to consider the colored  butterflies  from the
perspective of thermodynamics.

\subsection{Gibbs phase rule}\label{gibbs}
The first and second laws of thermodynamics constrain the shape of
phase diagrams. The phase rules depend on the choice of the
independent thermodynamic coordinates $X$ be they extensive, such
as $X=(E,V,N)$, or intensive, such as $(P,T)$.

Let $X=(E,V,N)$ be the extensive coordinates of a simple
thermodynamic system. $X$ and $\lambda X$ with $\lambda >0$ are
thermodynamically equivalent systems, while $X$ and $Y\neq \lambda
X$ are not. Mixing $X$ and $Y$, in any proportion, is, in general,
an irreversible process. The second law then says that the entropy
of the mixed system is not smaller than the sum of the entropies
of its constituents. Namely, for $0\le \lambda\le 1$
\begin{equation}\label{eq:convexity}
S(\lambda X +\lambda ' Y)\ge S(\lambda X)+S(\lambda 'Y)=\lambda S(
X)+\lambda 'S(Y), \quad \lambda '=1-\lambda
\end{equation}
The first law, conservation of energy, (plus conservation of
number of particles and additivity of volumes), was used in the
first step and the extensivity of the entropy, $S(\lambda
X)=\lambda S(X)$, in the second. Eq.~(\ref{eq:convexity}) says
that the entropy $S(X)$ is a concave function of its arguments.
This embodies the basic laws of thermodynamics.

Equality  in Eq.~(\ref{eq:convexity}) holds if mixing is
reversible which is, of course, the case if a phase is mixed with
itself. It is also the case if coexisting phases are mixed:
Clearly one can separate ice from water by mechanical means alone.
The geometric expression of equality in Eq.~(\ref{eq:convexity})
is that $S$ contains linear segment: For a pure phase this is the
half line $S(\lambda X)=\lambda S(X)$. When $X\neq \lambda Y$, are
in coexistence   $S$ contains a two dimensional cone: $S(\lambda_1
X+\lambda_2 Y)=\lambda_1S (X)+\lambda_2S (Y), \ \lambda_{12}>0$.
(This notion extends to multiple phase coexistence.)


Positivity of the temperature implies that $S$ is an increasing
function of $E$. Consequently, $S(E,V,N)$ can be inverted to give
the internal energy $E(S,V,N)$. Since $S$ is a concave function of
its arguments $E(S,V,N)$ is a convex function of its arguments
(which are the extensive state variables). Its Legendre transform
with respect to all its arguments, gives a function of the
intensive variables $T,P$ and $\mu$ alone which, by scaling, must
be identically zero. This is the Gibbs-Duhamel relation. It
determines the pressure $P$ as a convex function of the remaining
intensive coordinates, $(T,\mu )$:
\begin{equation}\label{eq:pressure}
PV=\mu N+TS-E
\end{equation}
The pressure is a convenient object to consider because all the
terms on the rhs of Eq.~(\ref{eq:pressure}) admit a simple
representation in statistical mechanics. $-P$ is sometimes called
the grand potential, e.g. \cite{ref:landau-sm}. Since the pressure
is the Legendre transform of the internal energy with respect to
$S$ and $N$. The convexity of $E$ then implies the convexity of
the pressure with respect to $T$ and $\mu$.

Now, it is a consequence of the duality of the Legendre transform
that if $E$ has a linear segment of length $\Delta X$ then its
Legendre transform $P$ has a corresponding jump in gradient with
$\Delta (\nabla P)=\Delta X$. It follows  that pure phases
correspond to points where $P(T,\mu)$ has a unique tangent, while
two phase coexist at those points $(T,\mu)$ where $P$ has two
(linearly independent) tangents planes. (Triple points are
similarly define.)

It is now a fact about convex functions that almost all points
have a unique tangent while the set with multiple tangents has
codimension $1$ (in the sense of comparing Hausdorff dimensions).
A geometric proof of this fact can be found in
\cite{ref:falconer}. This gives a weak version of the Gibbs phase
rule: If one considers the pressure $P$ as function of $(T,\mu)$,
(or alternatively, chemical potential $\mu$ as function of
$(P,t)$,) then pure phases are the typical sets while phases
coexist on exceptional, (i.e. small), sets.
\subsection{Magnetic systems}

At $T=0$ the entropy term in Eq.~(\ref{eq:pressure}) drops. For a
system of non-interacting Fermions  all single particle states
below $\mu$ are occupied, while those above are empty. This says
that for the single particle Hamiltonian $H$ and area $A$ the
pressure  is
\begin{equation}\label{eq:potential}
P=\lim_{A\to\infty}\ \frac 1 A\,Tr\,
\bigg(\big(\mu-H\big)_+\chi(A)\bigg)
\end{equation}
$\chi$ is the characteristic function of the area and
$x_+=x\,\theta(x)$ with $\theta$ a  unit step function.

Let $B$ denote the magnetic induction (i.e. the macroscopic
average of the local magnetic field \cite{ref:landau}). The
Hamiltonian is a function of $B$ and so is the pressure. The
density $\rho$ and the (specific) magnetization $M$ are then given
by \cite{ref:landau}
\begin{equation}\label{eq:N&M}
\rho=\Partial{P}{\mu},\quad M=\Partial{P}{B}\ .
\end{equation}
The Hall conductance is  thermodynamically defined by
\begin{equation}\label{eq:hall}
\sigma_H=\Partial{\rho}{B}=\Partial{M}{\mu}\ .
\end{equation}
It follows that, in the the wings of the butterflies where the
Hall conductance is quantized $P$ is given by:
\begin{equation}\label{eq:pingap}
P(\mu,B) =\sigma_g B(\mu-\mu_g) \,,
\end{equation}
where $g$ is a discrete wing label. The wings represent pure
phases since $P$ has a unique tangent in the gaps.


\subsection{The order of the transitions}
$P$, $ \rho$ and $M$ are bi-linear in $\mu$ and $B$  in the gaps.
$P$ and $\rho$ are actually also continuous functions of $\mu$ on
the spectrum. For rational flux this is a consequence of Floquet
theory. For irrational flux this can be seen by a limiting
argument.

 At the same time,  the Hall
conductance, being integer valued on a set of full measure, can
not be extended to a continuous function \footnote{ The
magnetization does not extend to a continuous function on the
spectrum for rational fluxes \cite{ref:gat}.}. (If fact, it is not
even bounded.) The continuity of the first derivative and the
discontinuity of the second derivatives, makes the phase
transitions in $\mu$ second order according to the Ehrenfest
classification \cite{ref:callen}.

\subsection{Phases and their boundaries}

In the colored Hofstadter butterflies  pure phases are open sets.
The boundary of a given phase, say the red wing, is a curve; It is
not a smooth curve as at rational values of $B$ is has distinct
tangents, but it is still a curve of Hausdorf dimensions one
\cite{ref:avron-osadchy}. This is reminiscent of the Gibbs phase
rule. Note, however, that the notion of the boundary of a pure
phase, and the notion of phase coexistence, are distinct.  The
phase with Hall conductance 1 meets the phase 0 at a single point,
at the tip of the butterfly, not on a line, as one might expect by
the Gibbs phase rule. This holds in general: The boundary of the
phases $i$ intersects the boundary of the phase $j$ on a set of
codimension 2, not 1 \cite{ref:avron-osadchy}. Moreover, any small
disc that contains two distinct phases of the butterfly contain
infinitely many other phases.
\subsection{Magnetic domains and phase coexistence}
Is the fractal phase diagram of the butterfly in conflict with
basic thermodynamic principles?

The Gibbs phase rule one finds in classical thermodynamics
\cite{ref:callen} says that two phases meet on a smooth curve
which is clearly not the case for the butterfly. However, this
strong version of Gibbs rule involves assumptions of smoothness of
free energies that may or may not hold. Convexity alone gives a
weaker version of the Gibbs phase rule, which we briefly discussed
in section \ref{gibbs}, which allows for all kind of wild
behaviors, and does not rule out fractal phase diagrams like the
butterfly.  \footnote{Instructive examples are given in p.8 of
\cite{ref:roberts}. I thank Aernout van Enter for pointing out
this example to me and for a clarifying discussion on the Gibbs
phase rule.}.

More worrisome is the lack of convexity of the pressure,
$P(\mu,B)$ which is manifest in the periodicity of
Fig.~\ref{fig:tb} in $B$. This raises the question if this
reflects a problem with the Hofstadter model. It does not. A
little reflection shows that rather, it a consequence of choosing
$B$, the magnetic induction, as the thermodynamic coordinate. In
the remaining part of this section I shall explain why it is
actually more natural to choose for the independent thermodynamic
variable  the magnetic field $H$ and the difficulties in drawing
the butterflies in the $\mu-H$ plane.


Imagine a two dimensional system with finite width which is broken
to domains. Assume that the magnetic field in each of the domains
is perpendicular to the plane and is constant through the given
domain. Since $\nabla\times H=0$, the magnetic field  $H$ must be
the same in adjacent domains.  Hence the notion of constant
magnetic field is constant $H$, while $B$ will not be constant if
the system breaks into domain. The problem with $H$ constant is
more difficult because it is $B$, not $H$, that enters in the
Hamiltonian \cite{ref:landau}.

Given the colored butterfly as function of $B$ what can one say
about the colored butterfly as a function of $H$? Recall that $B$,
$H$ and the magnetization $M$ are related by
\begin{equation}\label{eq:BHM}
  B=H+4\pi M
\end{equation}
Since $M$ is a function of $B$ so is $H$. However, $B$ may fail to
be a (univalued) function of $H$. This is the case if $-4\pi
\partial_BM\ge 1$; If the magnetic susceptibility is sufficiently negative.
When this happens, the relation $H(B)$ can not be inverted to a
function $B(H)$. Domains with different values of $M$ and $B$ may
then form and coexist \cite{ref:landau,ref:condon}.

The condition for coexistence is a stability condition: The system
will pick a value of $B$, consistent with $H$, that will minimize
the entropy. However, at $T=0$ the entropy of a gas of Fermions
vanishes, so the different solutions $B_j(H,\mu )$ all give the
same entropy, zero. This suggests that all the $B_j$ represent
phases at coexistence.

There is no reason why this degeneracy will hold if $T$ is not
strictly 0. Then, for most values of $H$ a distinguished solution
of $B_0(H,\mu )$ will be picked. The simple scenario is that that
$B_0(H,\mu )$ will depend, for most $H$, continuously on $H$. In
these intervals, the phases of the colored butterfly in $(\mu,H)$
will be a deformed version of the phases in $(\mu,B)$. However,
since a values of $B_0(H, \mu )$ is picked by a minimization
procedure, there is no guarantee for continuity and  $B_0(H,\mu )$
will be, in general, a discontinuous function of its arguments. At
the discontinuities, major qualitative changes in the diagram will
take place and it is interesting to investigate the colored
butterfly in the $\mu-H$ plane.

Another open problem in this context is to analyze the domain
structure for coexisting phases. The quasi-periodic character of
the electronic problem for irrational fluxes suggest that the
domain structure could be rich and interesting as well, (e.g.  a
quasi-periodic domain structure for irrational fluxes).

\section{Duality}
We now turn to the duality relating the two diagrams.
\subsection{Weak magnetic fields}

Consider the ``Bloch band'' dispersion relation
\begin{equation}\label{eq:dispersion}
\epsilon(k)= \cos (k\cdot a) +\cos(k\cdot b)
\end{equation} on the two dimensional Brillouin zone.
$a,b$ are the unit lattice vectors. The Hamiltonian describing a
weak external magnetic field  is obtained by imposing the
canonical commutation relation
\begin{equation}\label{eq:commutation}
[k\cdot a,k\cdot b]=i a\times b\cdot B\ =i\,{\Phi}.
\end{equation}
This procedure is known as the Peierls substitution
\cite{ref:spohn}.  The model is known as the Harper model, after a
student of Peierls.  The spectrum, plotted in fig.~\ref{fig:hof},
is a set of measure zero and so invisible in fig.~\ref{fig:tb}.
Figs. \ref{fig:tb},\ref{fig:hof} describe disjoint and
complementary sets, whose union is the plane.

Although there is considerable interest in Hofstadter model for
its own sake (see e.g. \cite{ref:avila} and references therein)
its physical significance to the two dimensional electron gas is
limited. One reason is that the flux $\Phi$, even for the
strongest available magnetic fields, is tiny and only a horizontal
sliver of the diagram in Fig.~\ref{fig:tb} near zero flux can be
realized. Moreover, $\Phi$ of order one is presumably outside the
region of weak field for which the model approximates the
Schr\"odinger equation.

By gauge invariance, time-reversal and electron-hole symmetry, the
pressure satisfies \cite{ref:gat}
\begin{equation}\label{eq:symmetry-omega}
P(\mu,\Phi)=P(\mu,-\Phi)=P(\mu,\Phi+1)=-\mu+P(-\mu,\Phi)
\end{equation}
This give Fig. \ref{fig:tb} its symmetry.

\subsection{Strong magnetic fields}

A classical charged  particle in homogeneous magnetic field moves
or a circle. The center of the circle is, classically,
\begin{equation}\label{eq:c-v}
c=x+\frac{v\times B}{B^2}
\end{equation}
$c$ commutes with $v$, but the  components of the center do not
commute, rather they satisfy the canonical commutation relations
\begin{equation}\label{eq:v}
[c\cdot a^*,c\cdot b^*]=-i\, \frac{B\cdot a^*\times
b^*}{B^2}=-i\frac {(2\pi)^2} {\Phi}\ .
\end{equation}
 $(a^*,b^*)$ are dual vectors to $(a,b)$.

If the wave function $\psi$ belongs to a given Landau level then
the shifts $e^{ic\cdot \alpha}\psi$, for $\alpha\in\mathbb{R}^2$,
span the spectral subspace of that level. This means that the
Hamiltonian
\begin{equation}\label{eq:sl}
\cos (c\cdot a^*) +\cos (c\cdot b^*),
\end{equation} acts within  Landau levels.
For large $B$ it approximates the  periodic potential $\cos
(x\cdot a^*) +\cos (x\cdot b^*)$, which couples different Landau
levels. This is seen from the fact that in a given Landau level
$v=O(\sqrt B)$ hence , by Eq.~(\ref{eq:c-v}), $c\approx x$ for
large $B$. The Hamiltonian in Eq.~(\ref{eq:v}) is the same as that
of Eq.~(\ref{eq:dispersion}), except that in the commutator
$\Phi/2\pi$ of Eq.~(\ref{eq:commutation}) is replaced by
$2\pi/\Phi$ of Eq.~(\ref{eq:v}).

Although the spectral problem of the two models is essentially the
same the phase diagrams are different. This is explained in the
next subsection.

Unlike the tight-binding model which is mostly of academic
interest, the model of a split Landau level can be realized in
artificial superlattices that accommodate a unit of quantum flux
at attainable fields.


\subsection{Thermodynamic duality}

The pressure of a split Landau level, $P_l$, and split Bloch band
$P_b$, for any temperature $T$, are related by \cite{ref:gat}
\begin{equation}\label{eq:omega-tb-ll}
P_l(T,\mu,\Phi/2\pi)=\frac {\Phi}{2\pi}\,P_b(T,\mu,2\pi/\Phi)
\end{equation}
This is a duality transformation: It is symmetric under the
interchange \hbox{$b\longleftrightarrow l$}. It implies that the
thermodynamics of the split Bloch band determine the
thermodynamics of a split Landau level and vice versa. The factor
$\Phi$ on the right is the reason that $P_l$ is not periodic,
although $P_b$ is.

A check on the factor $\Phi/2\pi$ comes by considering large
$\mu$. Then, the tight binding model has all sites occupied and
the electron density is $\rho_b \to 1$. This implies  $P_b\to
\mu$. In contrast, a full Landau level,  has electron density that
is proportional to the flux through unit area: $\rho_l\to
\Phi/2\pi$ so $P_l\to \Phi \mu/2\pi$.

The magnetization and the Hall conductances of the two models are
therefore related by:
\begin{eqnarray}\label{eq:m*-to-m}
m_l(\mu,T,2\pi/\Phi)&=&-\,\frac 1 {2\pi}\,P_b(\mu,T,\Phi/2\pi)-\frac{\Phi}{2\pi}
\,m_b(\mu,T,\Phi/2\pi);
\nonumber \\
\sigma_l(\mu,T,2\pi/\Phi)&=&\,\frac 1
{2\pi}\,\rho_b(\mu,T,\Phi/2\pi)-\frac{\Phi}{2\pi}\,\sigma_b(\mu,T,\Phi/2\pi)
\end{eqnarray}
When $\mu$ is large, $\sigma_b=0$, since a full band is an
insulator. At the same time, a full Landau level has a unit of
quantum conductance, $\sigma_l=1/2\pi$, in agreement with the
Eq.~(\ref{eq:m*-to-m}).

{\bf Acknowledgment: } I thank A. van Enter and O. Gat for useful
discussions and criticism. This work is supported by the Technion
fund for promotion of research and EU grant HPRN-CT-2002-00277.
\section{Appendix: Diophantine equation} Let me finally describe the
algorithm of \cite{ref:tknn} for coloring the gaps in the
butterfly fig.~\ref{fig:tb}. Suppose that the magnetic flux
through a unit cell is $\frac p q$. For $p$ and $q$ relatively
prime, define the conjugate pair $(m,n)$ as the solutions of
\begin{equation}\label{D}
pm-qn=1
\end{equation}
$m$ is determined by this equation modulo $q$ and $n$ modulo $p$.
The algorithm for solving Eq.~(\ref{D}) is the division algorithm
of Euclid. (Standard computer packages for  finding the greatest
common divisor of $p$ and $q$, yield also $m$ and $n$ such that
$pm+qn=\gcd(p,q)$.) The Hall conductance $k_j$, associated with
the j-th gap, in the tight binding case, is given by
\cite{ref:tknn}
\begin{equation}\label{DO}
k_j= j m \ mod \ q, \quad | k_j|\le q/2
\end{equation}
In the case of  split Landau band,  Eq.~(\ref{DO}) again
determines $k_j$ provided $p$ and $q$ are interchanged.

\clearpage
\begin{figure}
\caption{Colored Hofstadter butterfly for Bloch electrons in weak
magnetic field. The horizontal axis is the chemical potential; the
vertical axis is the magnetic flux through the unit cell. The
diagram is periodic in the flux and one period is shown.  It
admits a thermodynamic interpretation of a phase diagram.}
\label{fig:tb}
\end{figure}
\begin{figure}
\caption{The original, monochrome, Hofstadter butterfly, shows the
spectrum, on the horizontal axis, as function of the flux $\Phi$
which is the vertical axis. The spectrum is the complement of the
colored set shown in fig. \ref{fig:tb}. } \label{fig:hof}
\end{figure}
\begin{figure}
\caption{$S(E,V,N)$ is a concave function shown here for $N$
fixed. The strictly convex pieces are associated with pure phases.
The ruled piece is where two phases coexists. The boundary of the
region of coexistence is shown as a black line.}
\label{fig:coex-flat}
\end{figure}
\begin{figure}
\caption{Colored Hofstadter butterfly for Landau level split by a
super-lattice periodic potential. The horizontal axis is the
chemical potential; the vertical axis is the average number of
unit cells associated with a unit of quantum flux. As the number
increase by one the pattern repeats but with a different coloring
codes.} \label{fig:sl}
\end{figure}

\end{document}